\documentclass[times, onecolumn,11pt]{IEEEtran}
\usepackage{amssymb}
\usepackage{tipa}
\usepackage{pifont}
\usepackage{amssymb}
\usepackage{stmaryrd}
 \usepackage{amsmath}

\usepackage{subfigure}
\usepackage{epsfig}
\usepackage{graphicx}
\usepackage{setspace}
\usepackage{colortbl}
\ifodd 0
\newcommand{\com}[1]{\textbf{\color{blue} (COMMENT: #1)}} 
\else

\newcommand{\com}[1]{}
\fi

\makeatletter
\def\ps@headings{%
\def\@oddhead{\mbox{}\scriptsize\rightmark \hfil \thepage}%
\def\@evenhead{\scriptsize\thepage \hfil \leftmark\mbox{}}%
\def\@oddfoot{}%
\def\@evenfoot{}}
\makeatother
\newcommand {\Real} {\mathbb {R}}
\newcommand {\Sources} {\mathcal {S}}
\newcommand {\Receivers} {\mathcal {R}}
\newcommand {\Complex} {\mathbb {C}}
\newcommand {\Integer} {\mathbb {Z}}

\newtheorem{theorem}{Theorem}

\newtheorem{lemma}[theorem]{Lemma}

\title{Achieving the Scaling Law of SNR-Monitoring in Dynamic Wireless Networks}

\author{Hongyi Yao$^{1}$,  Xiaohang Li$^{2}$, Soung Chang Liew$^{3}$\\
$^{1}$ California Institute of Technology $\;$
 $^{2}$ Purdue University   $^{3}$ The$\;$Chinese University of Hong Kong
}

\begin{document}

\maketitle

\begin{abstract}
The characteristics of wireless communication channels may vary with time due to fading,
environmental changes and movement of mobile wireless devices. Tracking and estimating channel gains
of wireless channels is therefore a fundamentally important element of many wireless communication
systems. In particular, the receivers in many wireless networks need to estimate the channel
gains by means of a training sequence. This paper studies the {\it scaling law} (on the network size) of the overhead for channel gain monitoring in
wireless networks. We first investigate the scenario in which a receiver needs to
track the channel gains with respect to multiple transmitters. To be concrete, suppose that there are $n$ transmitters, and that in the current round of channel-gain estimation,  $k\leq n$ channels suffer significant variations since the last round.
We prove that ``$\Theta(k\log((n+1)/k))$ time slots'' is the {\it minimum} overhead needed to catch up with the $k$ varied channels. Here
a time slot equals one symbol duration.
At the same time, we propose a {\it novel} channel-gain monitoring scheme named
ADMOT to achieve the overhead {\it lower-bound}. ADMOT leverages recent advances in compressive sensing in signal processing and interference
processing in wireless communication, to enable the receiver to
estimate all $n$ channels in a {\it reliable} and {\it computationally efficient} manner within $\mathcal O(k\log((n+1)/k))$ time slots. To our best knowledge,  all previous channel-tracking schemes require $\Theta(n)$ time slots regardless of $k$.
Note that based on above results for single receiver scenario, the scaling law of general setting is achieved in which there are {\it multiple} transmitters, relay nodes and receivers.

\end{abstract}

\section{Introduction}
\label{sec:Intro}
The knowledge of  channel gains is often needed in the  design of high
performance communication schemes~\cite{EffectThrouput,FoundWireless,InterferenceAware,Physicallayernetworkcoding,ANC}.
In practice, due to fading, transmit power instability,
environmental changes and movement of mobile wireless devices, the channel gains vary with time.
Tracking and estimating channel gains
of wireless channels is therefore fundamentally important~\cite{DistributedMonitorArcitecture,Colocatedwirelessmonitoring,MLLEEstimationMAC,MIMONoisyEst,ChannelEstiMultiUserDetect,CooperSynforChannelEstimation,EstimatMIMO-OFDM}.

An issue of interest is how to reduce the overhead of channel-gain estimation. On the one hand, if between two rounds of channel-gain estimation, the channels have varied significantly, then communication reliability will be jeopardized~\cite{Physicallayernetworkcoding,ANC,MIMONoisyEst}. On the other hand, if the frequency of channel-gain estimation is high, the overhead will also be high~\cite{EffectThrouput,DistributedMonitorArcitecture,AdaptiveCommunication}. Our approach is predicated on reducing the overhead in each round, while maintaining high accuracy.

We first consider the case in which a receiver needs to estimate the channel gains from $n$ transmitters~\cite{EffectThrouput,FoundWireless}. As a mental picture, the reader could imagine the receiver to be a base station, and the transmitters to be mobile devices.  To achieve reliable bit-error-rate (BER),  the frequency of estimation should be high enough~\cite{EffectThrouput}. Then it is likely that only a few of the $n$ channels have suffered appreciable changes since the last estimation. We make use of the techniques of compressive sensing and interference signal processing to reduce the time needed to perform the estimation in each round. We answer the following question:

Suppose that in the current round, there are at most $k \leq n$ channels suffering from appreciable channel gain variations. Given a target reliability for channel-gain estimation, what is the minimum overhead needed?

We answer this question by analysis and construction. We prove that the minimum number time slots needed for estimation is $\Theta(k\log((n+1)/k))$, and we propose a scheme (named ADMOT) that uses $\mathcal O(k\log((n+1)/k))$ time slots\footnote{ Note that $\mathcal O(k\log((n+1)/k))=\mathcal O(k\log(n/k))$. In the paper we
use $\mathcal O(k\log((n+1)/k))$ to avoid the confusing case where $k=n$.}. Note that in each time slot, every transmitter transmits one symbol. Thus, one time slot is also one {\it symbol duration}.

Note that the general network scenario is also studied in which there are {\it multiple} transmitters, relay nodes and receivers. Again, the scaling law of estimating all network channels is achieved in a {\it reliable}, {\it computational efficient} and {\it distributed} manner.

\subsection{Illustrating Example and Background Ideas}
\label{sub:illustrative}

\begin{figure}[htp]
  \begin{center}
\includegraphics[height=30mm,width=78mm]{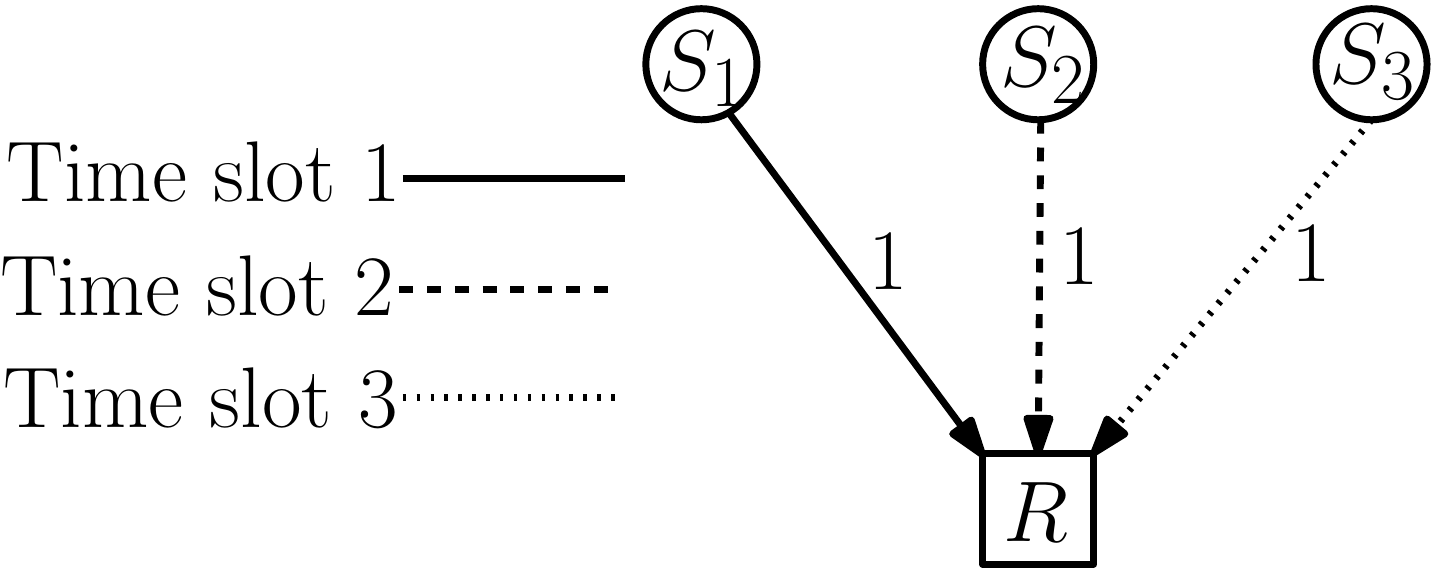}
 \end{center}
\caption{ The monitoring scheme based on scheduling.
The ``solid-line'', ``dashed-line'' and ``dotted-line'' are for the transmission of time slots 1, 2 and 3, respectively.}
\label{Fig:ToyNet}
\end{figure}

Consider a toy network consisting of three transmitting nodes $\{S_1,S_2,S_3\}$ and one receiving node $R$. The
 three channels $(S_1, R)$, $(S_2,R)$ and $(S_3,R)$ need to be estimated.
Without loss of generality, let all the initial channel gains of the three channels be $1$, and suppose one of the channel gains
changes to $x$  in the current time. The goal of monitoring is to identify the updated channel and the value of $x$.
A simplistic  monitoring scheme is to schedule transmissions on different channels in different time slots, as shown in Figure~\ref{Fig:ToyNet}.
In time slots 1, 2, and 3, sender $S_i$, $i=1, 2, 3$, sends probe signal $1$ to node $R$, respectively, so that $R$ can estimate the channel gain of $(S_i,R)$.  Thus, altogether three time slots are needed.

\begin{figure}[htp]
  \begin{center}
    \subfigure {\includegraphics[height=30mm,width=50mm]{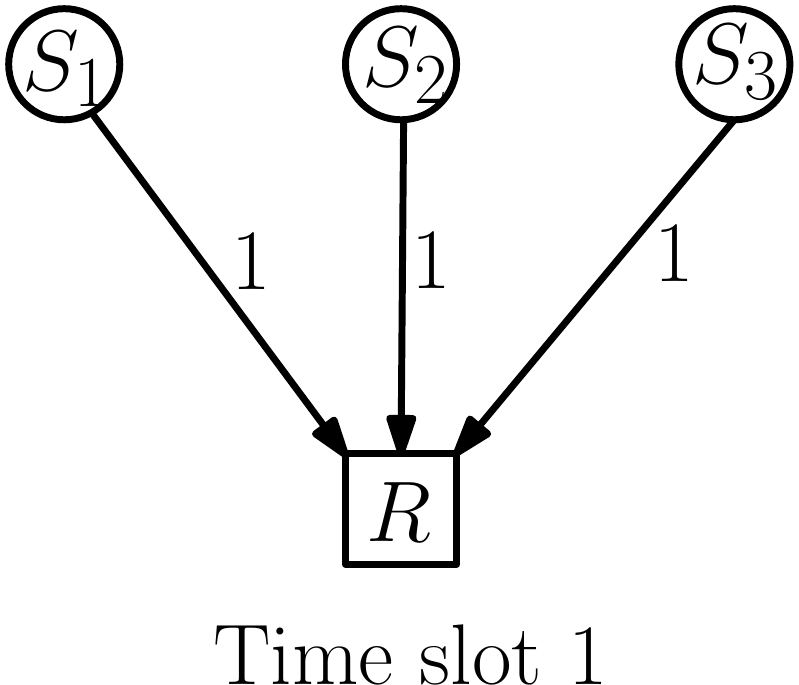}}\label{Fig:ToyNetInter1} \hspace{3mm}
    \subfigure {\includegraphics[height=30mm,width=50mm]{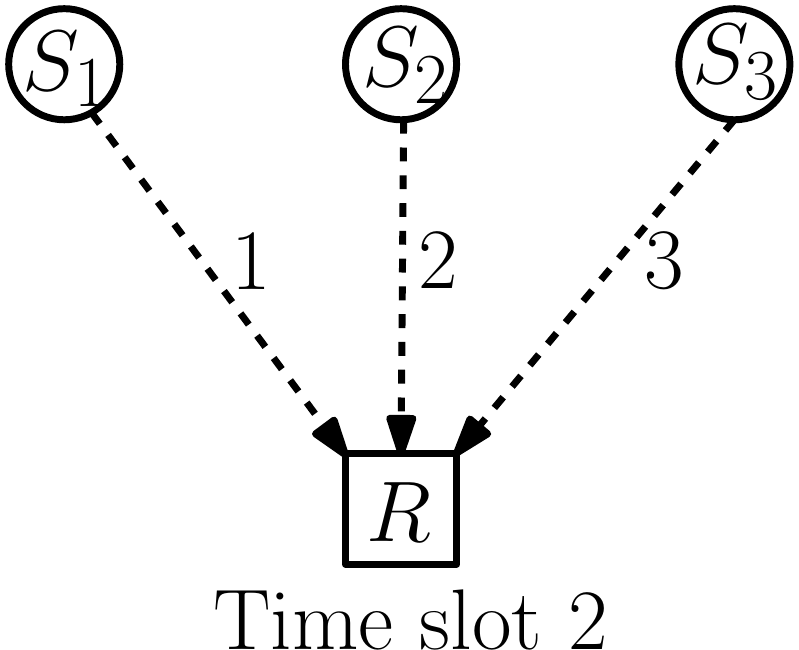}} \label{Fig:ToyNetInter2}\\
  \end{center}
  \caption{A better monitoring scheme. The first and second sub-figure show the transmissions in time slots 1 and 2, respectively.}
  \label{Fig:ToyNetInter}
\end{figure}

However, using the algebraic approach to exploit the nature of wireless medium, {\bf two} time slots are enough.
As shown in Figure~\ref{Fig:ToyNetInter}, in the first time slot $S_1$ and $S_2$ and $S_3$ all send $1$ to node $R$.
These three signals ``collide'' in the air,  but the collided signals turn out to be useful for our estimation. Let
 $y[1]$ denote the signal received by $R$ in the first time slot. We have $y[1]=3+(x-1)$.
In the second time slot $S_1$, $S_2$ and $S_3$ send $1$, $2$ and $3$, respectively. Thus, the received signal is $y[2]=6+i(x-1)$
if $(S_i,R)$ is the updated channel. At the end of the second time slot, $R$ computes $[y(1),y(2)]-[3,6]=(x-1)[1,i]$.
Since $[1,1]$ and $[1,2]$ and $[1,3]$ are mutually independent, $R$ can uniquely decode $i$ and $x$.

We summarize the main ingredients that give rise to the above saving as follows:

\noindent {\bf (a) Embracing Interference for Group Probing}. In group probing,  all the $n$ channels are probed simultaneously in each time slot. This is essential to get rid of the $\Theta(n)$ overhead in the traditional unit probing in which the channels are probed one by one.  Note the the $\Theta(n)$ overhead in unit probing is fundamental even if the number of channels suffering appreciable variations, $k$, is much smaller than $n$. This is because we do not know which channels have changed.

\noindent {\bf (b) Algebraic Distinguishability}. With respect to the illustrating example in Figure~\ref{Fig:ToyNetInter},  for each
 $i\in \{1,2,3\}$, the training data of $S_i$ ({\it i.e.}, $\{1,i\}$) induces an ``{\it algebraic fingerprint}'' for channel $(S_i,R)$. Due to the linear independence of the fingerprints, the one corresponding to
  the varied channel is not erasable  even under wireless interference. Note that it is not necessary to construct independent fingerprints by increasing the probing power. Later Section~\ref{Sec:UnitMonitor} shows that probing data with uniform magnitude but random signs suffice.

\subsection{Overview of Our Results}
\label{sub:overview}
For the scenario where a receiver wants to monitor the channel gains from $n$ transmitters, we first prove that the lower-bound of  overhead is $\Theta(k\log((n+1)/k))$ time slots, where $k\leq n$ is the number of channels suffering appreciable channel gain variations since the last round estimation. Note that when $k=n$, the overhead lower-bound is  $\mathcal O(n)$.

\begin{figure}[htp]
  \begin{center}
\includegraphics[height=47mm,width=98mm]{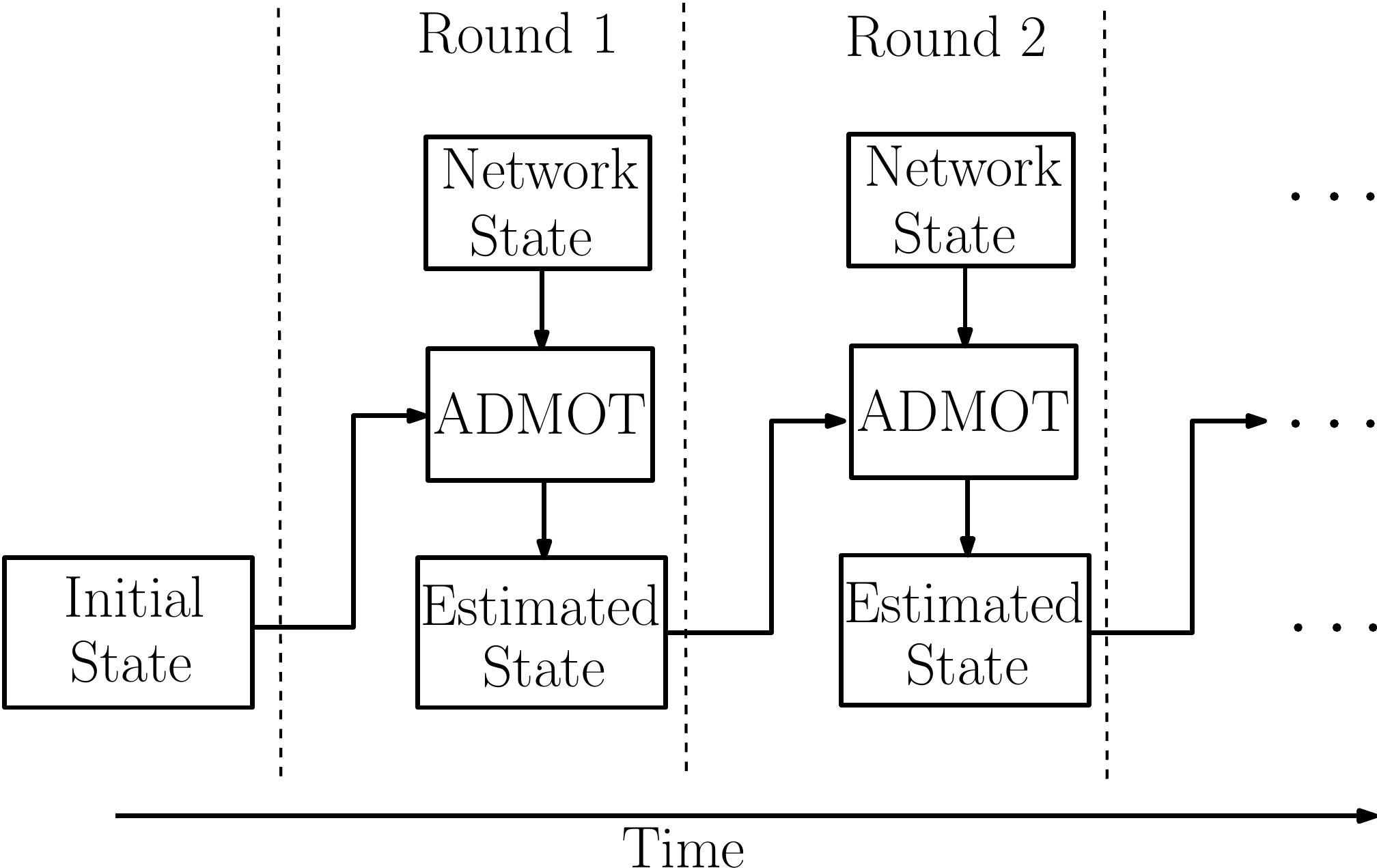}
 \end{center}
\caption{Systematical implementation of ADMOT.}
\label{Fig:SysADMOT}
\end{figure}

We then comprehensively develop the ideas in the above toy example and propose a scheme called Algebraic Differential SNR-Monitoring (ADMOT), which allows the receiver in the network to {\it reliably} monitor the channels with {\it minimum}
overhead. For a systematical view, Figure~\ref{Fig:SysADMOT} shows ADMOT in action for successive rounds\footnote{The interval between
two monitoring rounds depends on the statistics of the channel coherence time and channel stability requirement.}. In this figure, the {\it network state} is the set of channel gains of all channels, and the initial state is one in which
 every channel has zero channel gain. For the current {\it round} of monitoring,  ADMOT estimates
 the network state from the estimation of the previous round. In the following, we summarize the desirable features of ADMOT.

\noindent{\bf 1)} ADMOT is {\it optimal}. For any $k\leq n$, ADMOT allows the receiver to {\it reliably}
estimate all $n$ channels within $\mathcal O(k\log((n+1)/k))$ time slots, which matches the lower bound $\Theta(k\log((n+1)/k))$.  To our best knowledge,
 all previous monitoring results require $\Theta(n)$ regardless of $k$. Thus, ADMOT significantly reduces the overhead (compared with
 previous schemes) for small $k$, and preserves optimal performance even when $k$ is close to $n$. These arguments are
 also verified by simulation in Section~\ref{sec:Exp}.

\noindent{\bf 2)} Under ADMOT, the computational complexity of the receiver is dominated by {\it convex optimization programming}, which can be computed in an {\it efficient} manner~\cite{ConvexOPT}.

\noindent{\bf 3)} ADMOT is a {\it feedback-free} monitoring scheme, {\it i.e.}, the receiver does not need to send feedback during the monitoring and
no centralized controller is assumed. Thus, the probing data can be incorporated as packet-header for practical packet transmissions~\cite{EffectThrouput,ANC,MIMONoisyEst}.

\noindent{\bf 4)} ADMOT supports different {\it modulations}~\cite{FoundWireless} at the physical layer. For example, BPSK could be used. That is, the sources can code
 the training data (of ADMOT) into binary symbols for BPSK modulation, such that both channel attenuation and phase term  can be estimated.

We note that the {\it single receiver scenario} is the {\it fundamental} setting for studying the scaling law of monitoring wireless network. In Section~\ref{sec:ADMOTGeneral}, the above results are applied to achieve the scaling law of general communication networks in which there are {\it multiple} transmitters, relay nodes and receivers.

\subsection{Related Work}
\label{sub:RelatedWorks}

Previous works fall into the following two categories.

\noindent{\bf (a) Channel monitoring in wireless networks}. The works~\cite{MLLEEstimationMAC,MIMONoisyEst,ChannelEstiMultiUserDetect,CooperSynforChannelEstimation,EstimatMIMO-OFDM} designed probing data and estimation algorithm for estimating channel gains, and the works~\cite{InterferenceEstimation,InterferenceAware} proposed schemes to estimate channel interference. In the first set of works (which are related to our work), interference has not been shown to be an advantage (compared with {\it nonoverlapping} probing signals by different transmitters), and the overhead achieved is $\Theta(n)$.  Note that in the domain of wireless {\it network coding communication}, the work~\cite{Physicallayernetworkcoding} was the first to show the advantage of interference, and later the work~\cite{ANC} proposed an amplify-and-forward relaying strategy  for easy implementation.

\noindent{\bf (b) Compressive sensing for channel estimation}. ADMOT proposed in this paper uses recent advances of compressive
sensing  developed for sparse signal recovering~\cite{CompressiveSensing1,CompressiveSensing2}. Compressive sensing was used to recover the sparse features of channels, say
channel's delay-Doppler sparsity \cite{CSDelayChannelSparsity, SparceChannelCS}, channel's sparse multipath structure \cite{CSMultipathChannel,SparceChannelCS}, sparse-user detection~\cite{RandomAccess,CSReducingFeedback,DownlinkScheduling} and channel's
sparse response \cite{CSSparseChannel}. When applying above schemes to estimate all the $n$ channels from the transmitters, the overhead is at least $\Theta(n)$. In contrast, ADMOT uses compressive sensing to handle all channels' {\it differential} information (embedded in the overlapped probing) simultaneously, and achieves optimal overhead $O(k\log((n+1)/k))$.

Note that some previous schemes mentioned above estimate the property of a wideband channel, in which the channel gain varies across the frequency within the channel bandwidth. In contrast, this paper investigates the scaling law (on the network size) of wireless network monitoring. For the sake of exposition, we focus on narrowband channels in which the channel gain is flat across the bandwidth of the channel. We believe that within the same scaling law complexity,  ADMOT can easily be generalized to OFDM systems \cite{FoundWireless}, in which information is carried across multiple narrowband channels.

\subsection{Organization of the paper}
The rest of this paper is organized as follows. Section~\ref{Preliminary} formulates the problem. The scaling law theorem and the construction of ADMOT are presented in Section~\ref{Sec:UnitMonitor}.
In Section~\ref{sec:BPSK-ADMOT}, ADMOT is implemented by BPSK modulation. In Section~\ref{sec:ADMOTGeneral}, monitoring in general communication networks is studied.  Experimental results are shown in
 Section~\ref{sec:Exp} to support the theoretical analysis of ADMOT.

\section{Problem Setting and Preliminaries}
\label{Preliminary}

\subsection{Notation Conventions and Preliminaries}
\label{subsec:Notation}
Let $\Integer$ be the set of all integers and $\Integer^+$ be the set of all positive integers.
Let ${\Real}$ be the set of all real numbers ({\it i.e.}, the real field). For any $a$ and $b$ in $\Integer^+$,
let $\Real^{a\times b}$  be the set of all matrices with dimensions $a\times b$ and
components chosen from $\Real$. For any matrix $M\in \Real^{a\times b}$ and
$i\in \{1,2,...,a\}$ and $j\in \{1,2,...,b\}$, let $M(i,j)$ be the
$(i,j)$'th component of $M$.

Let $\Real^{a}$ be the set of all vectors with length $a$ and
components chosen from $\Real$.
For any vector $V\in \Real^{a}$ and
$i\in \{1,2,...,a\}$, let $V(i)$ be the
$(i)$'th component of $V$. {\it  Vectors in the paper are in the column form.}

For any vector $V\in \Real^n$, let $||V||_1=\sum_{i=1}^n |V(i)|$ denote the $\ell_1$-norm
and $||V||_2=\sqrt{\sum_{i=1}^n |V(i)|^2}$ denote the $\ell_2$-norm. For any vector $V\in \Real^n$ and non-negative
integer $k\leq n$, we define
the ``distance'' between $V$ and {\it $k$-sparsity}   by:
\begin{equation}
d_k(V)=||V-V^k||_1,
\label{eq:SparsityDistance}
\end{equation}
where $V^k$ is $V$ with all but
the largest $k$ components set to $0$. Vector $V$ is said to be {\it $k$-sparse} if and only if $d_k(V)=0$, that is
there are at most $k$ nonzero components in $V$.

Let $\Complex$ be the set of all complex numbers. The matrices and vectors over $\Complex$ has the similar definitions.
For any scalar, vector or matrix $X$ over $\Complex$, let $Re(X)$ be the real part of $X$ and $Im(X)$ be the imaginary part of $X$.
For vector $H\in \Complex^n$, let $||V||_2^2=||Re(H)||_2^2+||Im(H)||_2^2$.

Let $\mathcal {N}(\mu,\sigma^2)$ denote the normal distribution over {\it real field}, where $\mu$ is the {\it mean} and $\sigma^2$ is the variance. Note that throughout the paper, the logarithm function $\log(.)$ is computed over base $2$, {\it i.e.}, $\log(.)=\log_2(.)$.

\subsection{Communication Model for Wireless Network}
\label{sub:networkModel}
Consider a network where $\Sources=\{S_1,S_2,...,S_n\}$ is the set of transmitting nodes. We first consider the scenario where there is only one receiving node $R$.

 We assume all transmissions are slotted and synchronized. In each time slot, each and every transmitter transmits one symbol. Thus, one time slot is also one symbol duration and  slotted synchronization is the same as symbol synchronization as well.
Assume each  $S_i\in \Sources$ transmits symbol $X_i[s]\in\Complex$ in time slot $s$.
Thus, for time slot $s$ the received signal at $R$ is
\begin{equation}
Y[s]=\sum_{i=1}^n H(i) X_i[s] +Z[s],
\label{eq:InterChannel}
\end{equation}
where $H(i)\in \Complex$ is the channel gain of $(S_i,R)$ and $Z[s]\in \Complex$ is the noise. Note that both $Re(Z[s])$ and
$Im(Z[s])$ are {\it identically and independently distributed} (i.i.d.)$\sim \mathcal N (0,1)$ across all time slots. The {\bf state} of $R$ is defined to be a vector $H\in \Complex^{n}$, whose $i$'th component is $H(i)$.

The transmit power and noise power are both {\it normalized} to be equal to $1$ and the amplitude of the channel gain
$|H(i)|$ is the square-root of the signal-to-noise ratio (SNR) of $(S_i,R)$.
For instance, for channel $(S_i,R)$, assume $G(i)$ is the ``true'' channel gain, $P$ is the transmit power and $\sigma^2$ is the noise power.
Therefore, SNR$(S_i,R)=P|G(i)|^2/\sigma^2$.
 After{ normalizing} the transmit power and noise power, the amplitude of channel gain is $|H(i)|=|G(i)|\sqrt{P}/\sigma=\sqrt{\mbox{SNR}(S_i,R)}$.

As noted in Section~\ref{sub:overview}, such {\it single receiver scenario} is the {basic} setting for understanding the scaling law of  wireless network monitoring. In Section~\ref{sec:ADMOTGeneral}, we consider communication networks with {\it multiple} transmitters, relay nodes and receivers.

\subsection{Variation in Wireless Network}
\label{sub:NetworkVari}
Wireless network conditions vary with time due to fading, transmit power instability,
environmental changes and movement of mobile wireless devices.
For the purpose of communication, network variation is {\it mathematically}
equivalent to the variation of the state $H$ (the definition of $H$ can be
found in Section~\ref{sub:networkModel}).
Let $\hat H\in \Complex^n$ be the {\it a priori} knowledge of previous state held by the receiving node $R$, and $H$ be the current state. The monitoring objective of $R$ is to estimate $H$ using $\hat H$ and the received probes.
{\bf Note} if $R$ has no {\it a priori} knowledge of the previous network state, $\hat H$ is set to be the zero vector in $\Complex^n$.
For $\epsilon>0$ and non-negative integer $k\leq n$, the difference $H-\hat H$ is said to be $(k,\epsilon)$-sparse if and only if
$$d_k(Re(H-\hat H))\leq \epsilon \mbox{~~~and~~~}d_k(Im(H-\hat H))\leq \epsilon,$$ where function $d_k(.)$ is defined in Equation~(\ref{eq:SparsityDistance}).

For the simplicity, when $\epsilon$ is small, ``$(k,\epsilon)$-sparse'' is also said to be ``approx-$k$-sparse''.
Thus, for ``approx-$k$-sparse''  variation $H-\hat H$, there are at most $k$ channels suffering from {\it significant} variations for the channel gains, while the variations of other channels are negligible. In the following section ADMOT is proposed to catching up with the the $k$ major varied channels with minimum overheads.

\section{Achieving the Scaling law by ADMOT}
\label{Sec:UnitMonitor}
In this section, a {\it novel} wireless network monitoring protocol ADMOT is proposed to achieve the scaling law shown in Theorem~\ref{thm:ADMOT}.
To reduce the overhead of wireless monitoring, ADMOT fully develops the motivations shown in
Section~\ref{sub:illustrative}.
Furthermore, ADMOT exploits recent advances in the field of {compressive sensing} (\cite{CompressiveSensing1,CompressiveSensing2}), such that its {\it correctness} and {\it optimality}  can be {\it theoretically} proved.
A systematical view of ADMOT for consecutive wireless network monitoring can be found in Figure~\ref{Fig:SysADMOT}.

\subsection{Training Data of ADMOT}
\label{subsec:TrainDataADMOT}

The training data of ADMOT is denoted by matrix $\Phi$ with dimensions $N\times n$. Here, $n$ is the number of transmitters in the network and
$N$ is the upper-bound of time slots used by ADMOT.
Each component $\Phi(s,i)$ is generated independently from $\{-1,1\}$ with equal probability, for all $s$ and all $i$.
The $i$-th column of matrix $\Phi$ is assumed to be known {\it a priori} to transmitter $S_i$, for all $i\in \{1,2,\ldots,n\}$.
The knowledge of $\Phi$ can be broadcast by $R$ in the network setting stage\footnote{
To avoid the  overhead of broadcasting $\Phi$, we can generate $\Phi$ by practical pseudorandom generators (such as AES~\cite{ModernCrypto}). To be concrete, the $i$'th column of $\Phi$ could be the output of AES(i). Thus, each node
in the network can compute $\Phi$ using AES. Note that since
ADMOT can be simulated within polynomial time, pseudo randomness suffices~\cite{ModernCrypto}.}.

The training
data of each $S_i\in \Sources$ ({\it i.e.}, the $i$'th column of $\Phi$) is in fact the ``algebraic fingerprint'' of channel $(S_i,R)$. As  the toy example shown in Figure~\ref{Fig:ToyNetInter}, these fingerprints are ``highly independent'' such that the varied channels would expose their fingerprints even under interference. In the next subsection, using a convex program, ADMOT can catch up with the exposed fingerprints in an efficient and reliable manner.

\subsection{Complete Construction of ADMOT}
\label{subsec:consADMOT}
Before the detailed construction of ADMOT, a {\it convex-optimization} problem is proposed which serves as a submodule for ADMOT.

\begin{itemize}
\item  {\bf ConvexOPT($A,Y,\sigma$)}. The input of  ConvexOPT($A,Y,\sigma)$ is $A\in \Real^{m\times n}$ and $Y\in \Real^m$ and $\sigma>0$.
The output of ConvexOPT($A,Y,\sigma$) is the solution $X^*\in \Real^n$ to the following problem:
\begin{equation}
\mbox{min~}||X||_1\mbox{~~~subject to~~~~~}||AX-Y||_2\leq \sigma.
\label{eq:CSOPTIMIZATION}
\end{equation}
\end{itemize}
Note that ConvexOPT($A,Y,\sigma$) is a {\it second-order cone programming}
and can be solved efficiently~\cite{ConvexOPT}.

Let $m\leq N$ be the system parameter denoting the number of time slots used by the current round of ADMOT. We construct:

\begin{itemize}
\item {\bf ADMOT$(\hat H, {\cal S}, R, m)$}.

\item variables Initialization: Vector $H^*\in \Complex^n$ is the estimation of $H$, which is initialized to be zero vector.
Vector $Y\in \Complex^m$ is initialized to be zero vector. Let $\Phi_m$ be the matrix consisting of the $1,2,...,m$'th
rows of $\Phi$.

\item Step A: For $s=1,2,...,m$, in the $s$'th time slot:\footnote{Note that the probing scheme of ADMOT looks like CDMA~\cite{FoundWireless}. For the clarification we note the difference between ADMOT and
CDMA as: 1) CDMA is for information data detection, while ADMOT is for channel estimation. 2) CDMA requires near-orthogonal code sequence for each transmitter. In ADMOT, since $m$ could be much less than $n$, the training data sent by each $S_i\in \Sources$ can be far from orthogonal. However, the combination of ADMOT and CDMA is an interesting direction for future research.}
\begin{itemize}
\item  For any $S_i\in {\cal S}$, $S_i$ sends $\Phi(s,i)$.
\item Node $R$ sets $Y(s)$ ({\it i.e.}, the $s$'th component of $Y$) to be the received sample in the time slot.
Thus, $Y(s)=\sum_{i=1}^n\Phi(s,i)H(i)+Z(s)$, where $Z(s)$ is the noise in the time slot (see Section~\ref{sub:networkModel} for details).
\end{itemize}

\item Step B: Node $R$ computes $D\in \Complex^{m}$ as $D=Y-\Phi_m \hat H$. Thus, $D=\Phi_m (H-\hat H)+Z$.

\item Step C: Node $R$ runs $\mbox{ConvexOPT($\Phi_m,Re(D),\sqrt{2m}$)}$ and $\mbox{ConvexOPT($\Phi_m,Im(D),\sqrt{2m}$)}.$ Let the solutions be denoted by $Re(\Delta^*)\in\Real^n$ and $Im(\Delta^*)\in\Real^n$, respectively.

\item Step D: Node $R$ estimates $H$ by $H^* =\hat H+\Delta^*$.

\item Step E: {\bf End} ADMOT($\hat H,\Sources,R$ ).

\end{itemize}
\vspace{2mm}

Thus, ADMOT can be performed in a feedback-free manner, {\it i.e.}, the receiver does not need to send feedback during the monitoring and
no centralized controller is assumed. Under ADMOT, the computational complexity of the receiver node $R$ is dominated by running the second-order cone program ConvexOPT, which can be
solved in an efficient manner~\cite{ConvexOPT}.

\subsection{Main theorem of the paper}
\label{sub:TheoryADMOT}
The {\it main theorem} of the paper is:

\begin{theorem} {\it ADMOT is optimal.}
\hspace{-6mm}
\begin{itemize}
\item {\bf Scaling law}. For any $k\leq n$, when $H-\hat H$ is $(k,\delta\sqrt{k})$-sparse, any monitoring scheme achieving estimation  error $||H^*-H||_2\leq\mathcal O(\delta)$ requires at least $\Theta(k\log((n+1)/k))$ time slots.

\item {\bf Achievability}. Let $k>0$ be the maximum integer satisfying $C_0 k\log((n+1)/k)\leq m$ for a constant $C_0$, and $\delta>0$ be the minimum real number such that $H-\hat H$ is $(k,\delta\sqrt{k})$-sparse. The estimation error of ADMOT satisfies $||H^*-H||_2\leq\sqrt{2}C_1\delta+2C_2$ with a probability $1-\mathcal O\Big(e^{-0.15m}\Big)$. Here, $C_0$, $C_1$ and $C_2$ are  constants defined in Appendix~\ref{appendix: CS}.
\end{itemize}
\label{thm:ADMOT}
\end{theorem}

The detailed proof for the theorem is in Appendix~\ref{appendix: Proof}. We have the following remarks regarding the theorem.

\noindent{\bf Remark 1}: When $H-\hat H$ is approx-$k$-sparse ({\it i.e.}, $\delta$ is small), the theorem shows that ADMOT reliably estimates $H$ within $\mathcal O(k\log((n+1)/k))$ time slots, which achieves the scaling law.

\noindent{\bf Remark 2:} Recall that we normalize both probe power and noise power for the clarity. As shown in Section~\ref{sub:networkModel}, the ``true'' channel gain $G(i)$ is in fact $G(i)=H(i)\sigma/\sqrt{P}$, where $P$ is the transmit power and $\sigma^2$ is the noise power. Thus, the ``true'' channel gains are estimated by $G^*=\sigma H^*/\sqrt{P}$. When $G-\hat G$ is $(k,\delta_G\sqrt{k})$-sparse, $H-\hat H$ is $(k,\delta\sqrt{k})$-sparse with $\delta=\delta_G\sqrt{P}/\sigma$. Thus, the estimation error of $G$
is then $||G^*-G||_2\leq (\sqrt{2}C_1\delta +2C_2)\sigma/\sqrt{P}=\sqrt{2}C_1\delta_G+2C_2\sigma/\sqrt{P}$. Thus, for large probing power $P$, the
 error term  $2C_2\sigma/\sqrt{P}$ caused by noise  disappears  and $||G^*-G||_2$ approximates $\sqrt{2}C_1\delta_G$.\footnote{Note that for the case where each node in $\Sources$ has different probing power, similar argument can be shown with somewhat unwieldy notations.}

\subsection{Adjusting system parameter $m$}
\label{sub:Adjustm}

The system parameter $m$ corresponds to the trade-off between  overheads and estimation errors. Ideally, we should choose $m=C_0 k\log((n+1)/k)$, where $k$ is the number of channel gains which suffer significant variations since the last round estimation. Thus, the receiver $R$ should estimate the {\it typical} number of varied channels  between two monitoring rounds, and then adjust $m$ for future rounds of ADMOT\footnote{Note that the receiver $R$ can inform its choosing of $m$ to other nodes by  broadcasting before the next round of ADMOT.
For instance, consider a cellular network where the receiver is the base station~\cite{EffectThrouput}. The information of $m$ can be delivered in the stage of {\it downlink} transmission.}.

This object can be achieved by analyzing the estimation error in the past rounds.
 To be concrete, consider  a past round of ADMOT with system parameter $m$. Let $Y$ be the received data in this round. Receiver $R$ divides $Y\in \Complex^m$ into two parts (one for ``estimation'' and the other for ``testing''): Vector $Y_1\in \Complex^{m-d}$ comprises of the first $m-d$ components of $Y$ and $Y_2\in \Complex^d$ comprises of the last $d$ components of $Y$.
Similarly, matrix $\Phi_{(m,1)}$ comprises of the first $m-d$ rows of $\Phi_m$ and $\Phi_{(m,2)}$ comprises of the last $d$ rows of $\Phi_m$.

Receiver $R$ runs Step B, C and D of ADMOT by using $Y_1$ and $\Phi_{(m,1)}$ (instead of $Y$ and $\Phi_m$, respectively).
Let $H^*_t$ be the estimation of $H$. Let $D_2=Y_2-\Phi_{(m,2)}H^*_t$ and $||H-H^*_t||_2=\varphi$. Then we have:

\begin{theorem}
\label{le:D2Prob}
The event $||D_2||_2^2>d(\varphi\sqrt{3/2}+2)^2$ happens with
a probability at most $O\Big(e^{-0.15d}\Big)$.
If $\varphi>2\sqrt{2}$, the event $||D_2||_2^2<d(\varphi/\sqrt{2}-2)^2$ happens with
a probability at most $O\Big(e^{-0.15d}\Big)$.
\end{theorem}

The proof is delivered into Appendix~\ref{appendix: EstK}.

The theorem shows that the estimation error $||H-H^*_t||$ preserves a close relationship with $||D_2||_2$.
Thus, when $||D_2||_2$ is large, $R$ concludes that  $m-d$ time slots do not suffice to estimate $H$. Note that since $d$ is relatively small (compared with $m$), $m$ time slots probing is also not reliable for estimating $H$. Thus, $R$ should {\bf increase} $m$ for future rounds of ADMOT.

On the other hand, small $||D_2||_2$ implies  $m-d$ time slots are sufficient for estimating $H$.
For precisely estimating the minimum time slots which suffice, $R$ can update $d$ to $2d$ and then re-computes $||D_2||_2$.

Then, $R$ repeats this process until it finds the minimum integer $p$ such that $m-pd$ time slots are not sufficient. In the end, based on $pd$, $R$ can choose an appropriate {\bf decreasing} of $m$ for future rounds of ADMOT.

\section{BPSK Implementation of ADMOT}
\label{sec:BPSK-ADMOT}

In this section, we describe ADMOT implementation using BPSK modulation~\cite{FoundWireless}. That is, we assume the symbols in $\Phi$ are BPSK symbols.

Each node $S_i\in \mathcal S$ transmits a BPSK symbol $\Phi(s,i)\in \{-1,1\}$   in  time slot $s$.
 All transmitters transmit their symbols on the angular carrier frequency $\omega$. Let $T$
  be the duration of a time slot. Thus, in continuous time, $S_i$  transmits the signal $x_i(t)=Re(\sum_s \Phi(s,i)p(t-sT)e^{j\omega t})$,
  where $p(t)=1$  for $0\leq t<T$; and $p(t)=0$ otherwise. Let the channel gain associated with $S_i$  be $H(i)=A_i e^{-j\theta_i}$,
  where $A_i\in [0,+\infty)$ is the amplitude of the channel gain and $\theta_i$ is the phase delay due to signal propagation delay.

The signal reaching $R$  from $S_i$  is then  $Re(\sum_s \Phi(s,i)p(t-sT)e^{j\omega t}A_i e^{-j\theta_i})$.
Taking into the consideration the signals from all nodes in $\Sources$ and the circuit noise, the combined signal at $R$ is
$$y(t)=\sum_{i} Re\Big(\sum_s \Phi(s,i)p(t-sT)e^{j\omega t}A_i e^{-j\theta_i}\Big)+z(t),$$
where $z(t)$ is the noise.

We assume $T\gg 1/\omega$. By matched filtering (i.e., multiplying $y(t)$  by $\cos(\omega t)$ and
integrating over successive symbol periods $T$; and multiplying $y(t)$  by $\sin(\omega t)$
and  integrating over successive symbol periods), we can get
\begin{eqnarray*}
& &Y_{cos}[s]=\sum_i A_i \cos(\theta_i) \Phi(s,i)+\frac{2}{T}\int_{(s-1)T}^{sT}z(t)\cos(\omega t)dt,\\
& &Y_{sin}[s]=\sum_i A_i \sin(\theta_i) \Phi(s,i)+\frac{2}{T}\int_{(s-1)T}^{sT}z(t)\sin(\omega t)dt.
\end{eqnarray*}

Note that the power of noise are normalized to $1$ such that
$\frac{2}{T}\int_{(s-1)T}^{sT}z(t)\cos(\omega t)dt$  and $\frac{2}{T}\int_{(s-1)T}^{sT}z(t)\sin(\omega t)dt$ are both i.i.d. $\sim\mathcal N(0,1)$.

Thus, there are two set of channels  and no inter-set inference happens. The state of the first set of channels is
$H_{cos}\in {\Real}^n$, where the $i$'th component is $H_{cos}(i)=A_i \cos(\theta_i)$. And state of the second set of channels is
$H_{sin}\in {\Real}^n$, where the $i$'th component is $H_{sin}(i)=A_i  \sin(\theta_i)$.

Following ADMOT, the receiver can estimate $H_{sin}$ and $H_{cos}$ simultaneously (as monitoring $Re(H)$ and $Im(H)$, see Section~\ref{subsec:consADMOT}).
Once $H_{sin}$ and $H_{cos}$ are estimated, $\{(A_i,\theta_i):S_i\in \Sources\}$ can be computed efficiently. To be concrete, let
$H_{sin}^*$ and $H_{cos}^*$ be the estimations of $H_{sin}$ and $H_{cos}$, respectively. Thus, $\theta_i$ can
be estimated by $\theta_i^*=\tan^{-1}(H_{sin}^*(i)/H_{cos}^*(i))$ and  $A_i$ can be estimated by $A_i^*=\sqrt{(H_{cos}^*(i))^2+(H_{sin}^*(i))^2}$.

Note that ADMOT can also be implemented with other modulations. Due to the limit of space, we only present the BPSK implementations for ADMOT in this paper.

\section{Achieving the Scaling Law for General Communication Network}
\label{sec:ADMOTGeneral}

In this section, we study the general communication network with multiple transmitters, receivers and intermediate relay nodes, e.g., the cooperative communication networks~\cite{CooperativeCommunication}.
To be concrete, we can  model the general communication network   as $(\Sources,\Receivers,\mathcal C)$, where  $\mathcal S=\{S_1,S_2,...,S_n\}$ is set of transmitters , $\Receivers=\{R_1,R_2,...,R_{n'}\}$ is the set of receivers and $\mathcal C=\{C_1,C_2,...,C_{n''}\}$ is the set of intermediate nodes for relaying. Let $\mathcal E$ denote the set of all the channels: $\{(S_i,R_j),(C_a,R_j),(S_i,C_a), (C_a,C_b):S_i\in \mathcal S, R_j\in \mathcal R, C_a,C_b\in \mathcal C\}$. Each channel in $\mathcal E$ is either used for communication or considered as a interfering channel,  and therefore needs to be monitored.

\begin{figure}[htp]
  \begin{center}
\includegraphics[height=35mm,width=68mm]{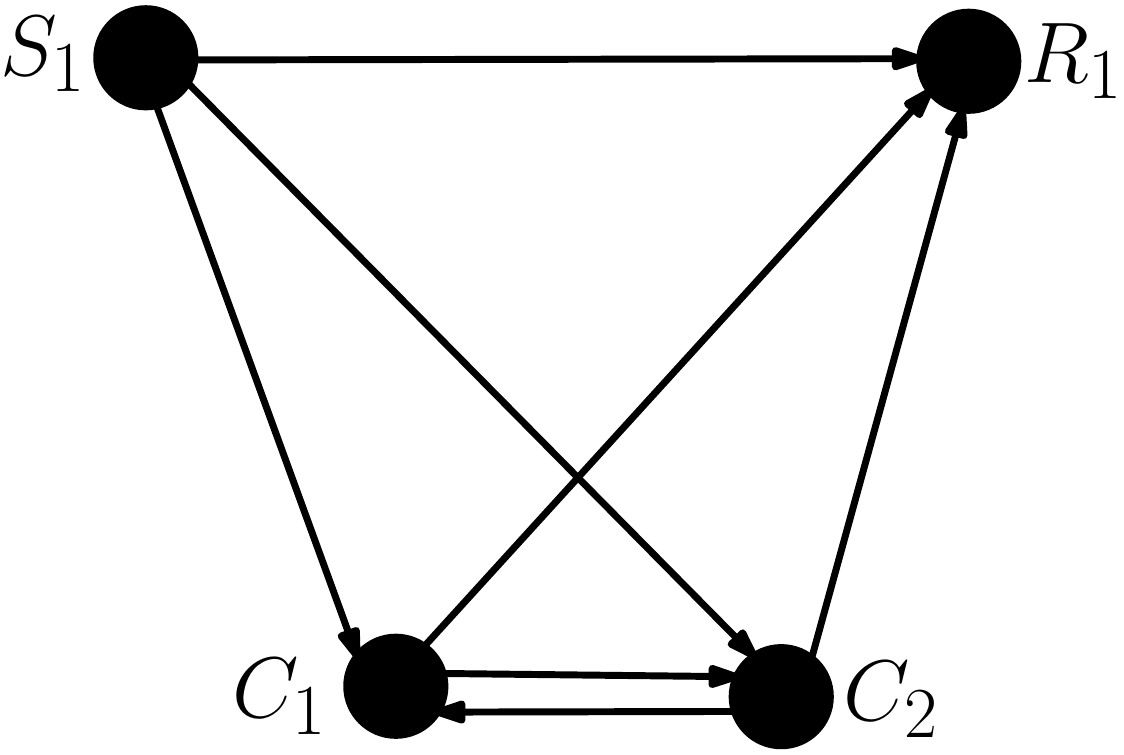}
 \end{center}
\caption{The channels in an illustrating communication network ($\Sources, \Receivers,\mathcal C$). In the figure, $\Sources=\{S_1\}$, $\Receivers=\{R_1\}$ and $\mathcal C=\{C_1,C_2\}$. The directed lines denote the channels which are used for communications and therefore require monitoring.}
\label{Fig:GeneralNet}
\end{figure}

For each node $\beta \in \Receivers\cup \mathcal C$, let $H_\beta\in \Complex^{n+n''}$ be the channel gains of the
channels from $\Sources\cup \mathcal C$ to $\beta$, and $\hat H_\beta\in \Complex^{n+n''}$ be the {\it a priori} knowledge of these channel gains preserved by $\beta$. Assuming for each node in $\Receivers\cup \mathcal C$, its state variation $H_\beta-\hat H_\beta$ is approx-$k$-sparse for $k\leq n+n''$. Using the scaling law in Theorem~\ref{thm:ADMOT}, at least $\Theta(k\log((n+n''+1)/k))$ time slots are needed.

Under the full-duplex model, where any node in $\mathcal C$ can transmit and receive in the same time slot, ADMOT$(\hat H_\beta, \Sources\cup\mathcal C, \beta, m)$ can be performed simultaneously for each $\beta \in \Receivers\cup \mathcal C$. Thus, we can achieve above overhead lower bound by choosing $m=\Theta(k\log((n+n''+1)/k))$.

For the half-duplex model, any node in $\mathcal C$ cannot transmit and receive in a same time slot.
Note that
in the half-duplex model, for any node $C_j \in \mathcal C$ the channel gain of $(C_j,C_j)$ is always assumed to be $0$.
In next subsection we propose a {\it non-straightforward} generalization of ADMOT (named ADMOT-GENERAL) to achieve the scaling law $\Theta(k\log((n+n''+1)/k))$.

Note that for both models, our achievable schemes can be performed in a distributed manner, {\it i.e.}, no centralized controller is needed.

\subsection{ADMOT-GENERAL}
To handle the half-duplex feature, for each time slot ADMOT-GENERAL randomly selects (with $1/2$ probability) nodes in $\mathcal C$ to send probe data, while other nodes
in $\mathcal C$ receive signal in the time slot.  To be concrete, let $\Phi\in \Real^{N\times (n+n'')}$ be the training data matrix. Each component of $\Phi$ is i.i.d. chosen from $\{0,-1,1\}$ with
a probability $\{1/2, 1/4, 1/4\}$. For each $S_i\in \Sources$, the $i$'th column of $\Phi$ is the training data
of $S_i$. For each $C_j\in \mathcal C$, the $n+j$'th column of $\Phi$ is the training data of $C_j$. In ADMOT-GENERAL, for the $s$'th time slot, if the training data of $C_j$ is zero, $C_j$ would receive signal in the time slot; Otherwise $C_j$ would send the corresponding probe data.

Choosing $m=3C_0' k \log (n/k)=\Theta{(k \log (n/k))}$ where $C_0'$ is a constant defined in Appendix~\ref{appendix: CS}, we have:

\begin{itemize}
\item {\bf ADMOT-GENERAL$(m)$}\footnote{For the simplicity, we omit other parameters $(\{\hat H_\beta: \beta \in \Sources \cup \mathcal C\}, \Sources, \mathcal C, \Receivers)$.}.

\item variables Initialization: For each $\beta\in \Receivers\cup \mathcal C$, vector $H_\beta^*\in \Complex^{n+n''}$ is the estimation of $H$, which is initialized to be zero vector.
For each $\beta\in \Receivers\cup \mathcal C$, let $\mathcal I_\beta=\{i: \Phi(i,n+j)= 0, i\in \{1,2,...,m\}\}$ if $\beta=C_j\in \mathcal C$, and $\mathcal I_\beta=\{1,2,...,m\}$ if $\beta \in \Receivers$. For each $\beta\in \Receivers\cup \mathcal C$, let $m_\beta=|\mathcal I_\beta|$, and $\Phi_\beta \in \Real^{m_\beta\times ({n+n''})} $ consist of the rows of $\Phi$ which are indexed by $\mathcal I_\beta$, vector $Y_\beta\in \Complex^{m_\beta}$ be initialized to be zero vector.

\item Step A: For $s=1,2,...,m$, in the $s$'th time slot:

\begin{itemize}
\item  For any $S_i\in {\cal S}$, $S_i$ sends $\Phi(s,i)$.

\item For any $C_j \in {\cal C}$, if $\Phi (s, n+j)=0$, $C_j$ receives signal in this time slot; Otherwise $C_j$ sends $\Phi_\beta(s_\beta,n+j)$, where $s_\beta\in \mathcal I_\beta$ is the index of the row in $\Phi_\beta$ which corresponds to
the $s$'th row of $\Phi$.

\item Any node in $\Receivers$ receives signal in this time slot.

\item For each node $\beta\in \Receivers\cup \mathcal C$, if $\beta$ received signal in this time slot, $\beta$ sets
$Y_\beta(s_\beta)$ to be the received sample.
Thus, $Y_\beta(s_\beta)=\sum_{i=1}^{n+n''}\Phi_\beta(s_\beta,i)H_\beta(i)+Z_\beta(s)$, where $Z_\beta(s)$ is the noise in the time slot.
\end{itemize}

\item Step B: For each $\beta\in \Receivers\cup \mathcal C$, $\beta$ computes $D_\beta\in \Complex^{m_\beta}$
as $D_\beta=Y_\beta-\Phi_\beta \hat H_\beta$. Thus, $D_\beta=\Phi_\beta (H_\beta-\hat H_\beta)+Z_\beta$, where $Z_\beta\in \Complex^m_\beta$
is the noise vector for $\beta$.

\item Step C: For each $\beta\in \Receivers\cup \mathcal C$, $\beta$ runs $\mbox{ConvexOPT($\Phi_\beta,Re(D_\beta),\sqrt{2m_\beta}$)}$ and $\mbox{ConvexOPT($\Phi_\beta,Im(D_\beta),\sqrt{2m_\beta}$)}.$\footnote{Note that if $\beta=C_j\in \mathcal C$, since the channel gain of $(C_j,C_j)$ is always assumed to be zero, the $n+j$'th components of $Re(D_\beta)$
    and $Im(D_\beta)$ are both fixed to be zero for running ConvexOPT. }
    Let the solutions be denoted by $Re(\Delta_\beta^*)\in\Real^{n+n''}$ and $Im(\Delta_\beta^*)\in\Real^{n+n''}$, respectively.

\item Step D: For each $\beta\in \Receivers\cup \mathcal C$, $\beta$ estimates $\beta$ by $H_\beta^* =\hat H_\beta+\Delta_\beta^*$.

\item Step E: {\bf End} ADMOT-GENERAL.

\end{itemize}
\vspace{2mm}

For each $\beta \in \Receivers \cup \mathcal C$, using Chernoff Bound~\cite{ProbabilityBook2005},  we have $m_\beta \geq m/3 = \geq C_0' k\log(n/k)$ with a probability at least $1-2^{-m}$. Using the achievability result in Theorem~\ref{thm:ADMOT}, we conclude $\beta$ can recover $H_\beta$ with bounded square root errors\footnote{Note that $\Phi_{\beta}$ satisfies RIP of order $k$ with a probability at least $1-2^{n}$ (see Appendix~\ref{appendix: CS}), which is the sufficient requirement for applying Theorem~\ref{thm:ADMOT} (see the proof in Appendix~\ref{appendix: Proof}).}.

\section{Performance evaluation}
\label{sec:Exp}

We evaluate ADMOT by implementing it in a systematical manner, as shown in Figure~\ref{Fig:SysADMOT}.
Let the $n=|\Sources|=500$, and the average channel SNR$=20$db.

Recall that a channel is said to preserve {\it stability} $x\%$ if the probability  is no more than $(1-x\%)$ that the channel suffers significant variation during the interval between two monitoring rounds.

In the simulation, let $H[r]$ the be state of the $r$'th round.
Thus, for the channel stability $x\%$, $H[r]= H[r-1]+\Delta[r]$, where $\Delta[r]$ is the variation. Each component of
 $\Delta[r]\in \Complex^n$, say $\Delta[r](i)$, is {\it independently} generated as: With a probability $x\%$, both $Re(\Delta[r](i))$ and
 $Im(\Delta[r](i))$ are uniformly chosen from $[-10,10]$; With a probability $1-x\%$, both $Re(\Delta[r](i))$ and
 $Im(\Delta[r](i))$ are uniformly chosen from $[-250,250]$.

We proceed ADMOT$( H^*[r-1],\Sources, R,m_r )$ for the $r$'th round estimating. Here, $ H^*[r-1]$
is the estimation of $H[r-1]$ in the $(r-1)$'th round, and the system parameter $m_r$ is chosen depending on the receiving data in the previous
rounds (see Section~\ref{sub:Adjustm} for details).

\begin{figure}[htp]
\begin{center}
\includegraphics[height=87mm,width=117mm]{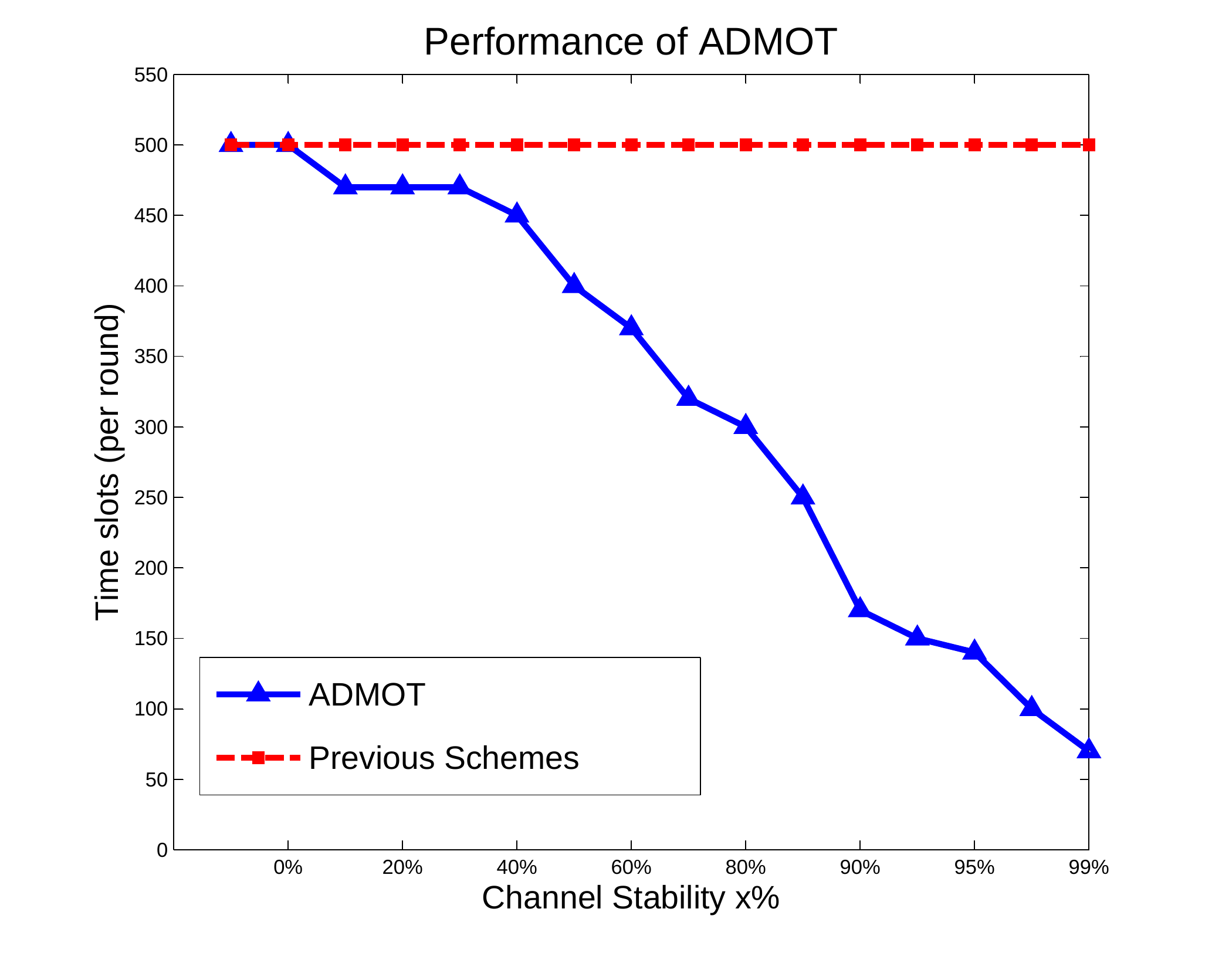}
 \end{center}
\caption{Comparison between ADMOT and previous monitoring schemes.}
\label{Fig:GlobalEstTime}
\end{figure}

Figure~\ref{Fig:GlobalEstTime} shows the average time slots (per round) used by ADMOT for $x\in(0,100)$. In the figure, the solid
line is for ADMOT, and the dashed line is for previous monitoring schemes (see related works in Section~\ref{sub:RelatedWorks}).
From the figure, we can see that ADMOT significantly reduces the overheads for the scenarios where $x$ is large, {\it  i.e.}, high channel stability is required. In the region where $x$ is small, ADMOT also preserves reliable performance.

We also provide the detailed simulations for the cases where channel preserves stabilities $80\%$, $90\%$, and $98\%$, respectively.

\begin{figure}[htp]
\begin{center}
\includegraphics[height=87mm,width=117mm]{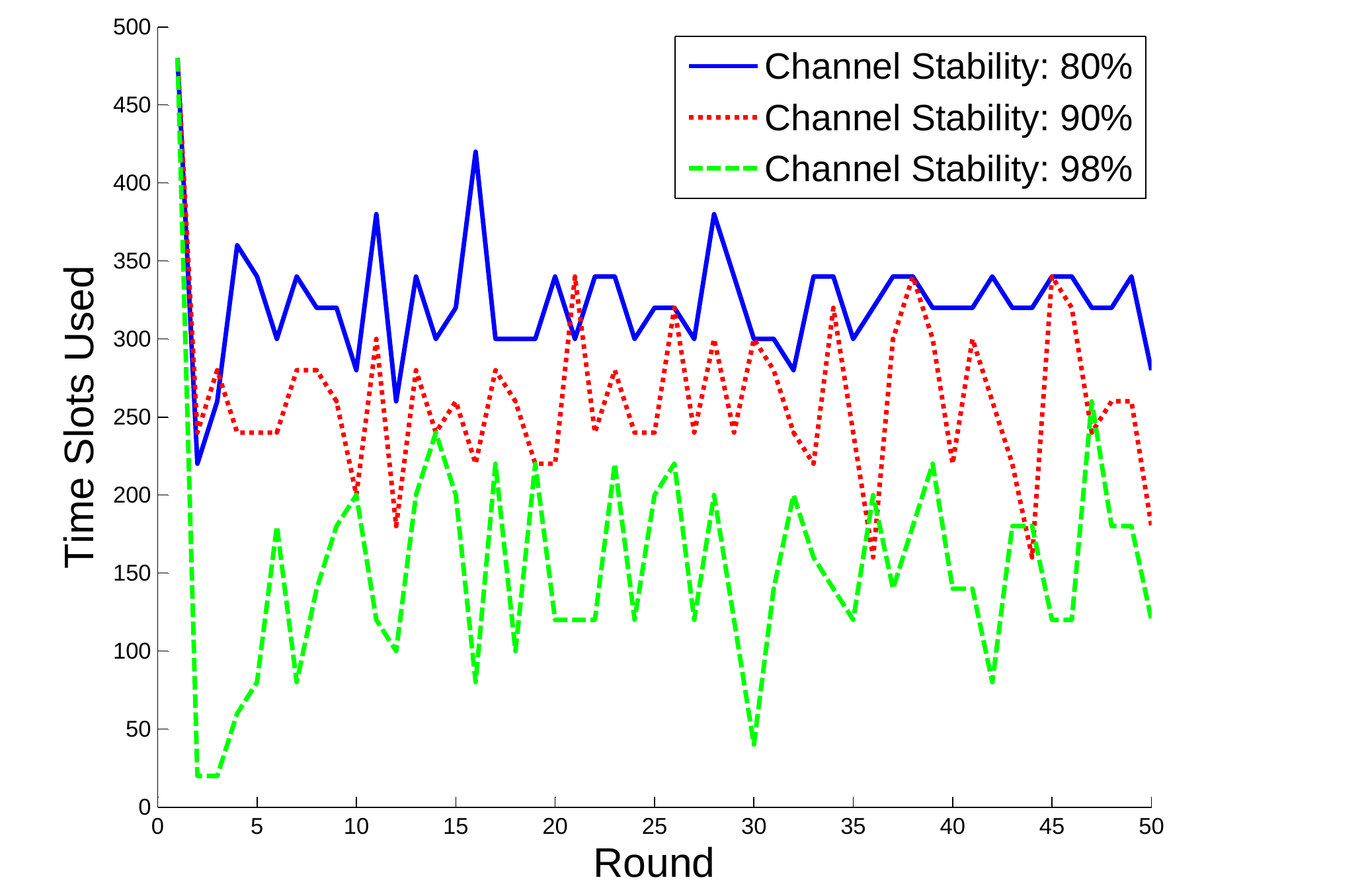}
 \end{center}
\caption{Estimating time for each stage of running ADMOT. }
\label{Fig:EstTime}
\end{figure}

Figure~\ref{Fig:EstTime} shows the time slots used by ADMOT for round $r\in \{1,2,..,50\}$. For the channel stability $80\%$, $90\%$, and $98\%$,
the average time slots used per round are $320$, $252$, and $140$, respectively. Note that since we assume zero knowledge for the initial
network state, the first round of each case costs almost $500$ time slots.

\begin{figure}[htp]
\begin{center}
\includegraphics[height=87mm,width=97mm]{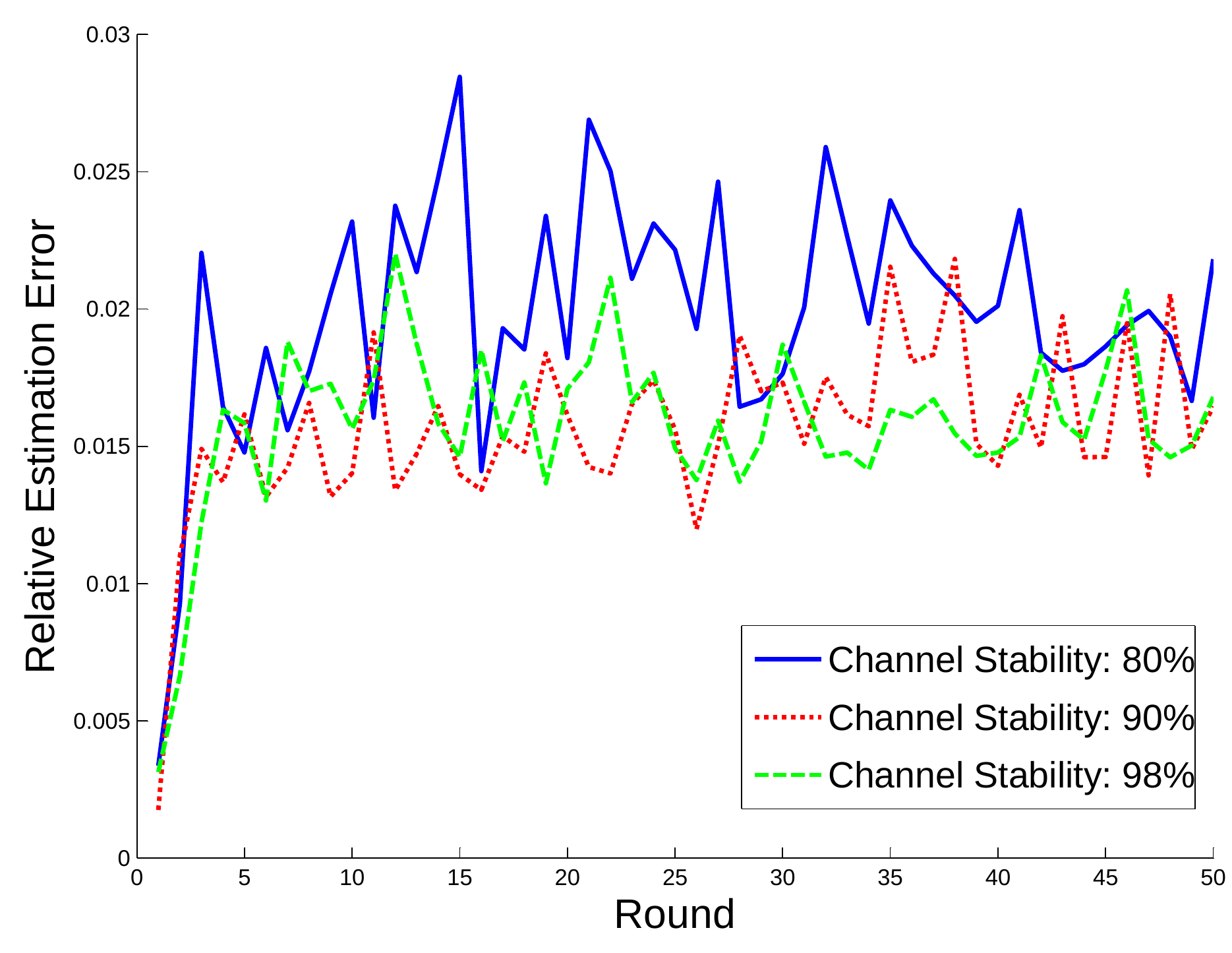}
 \end{center}
\caption{Relative estimating error for each stage of running ADMOT.}
\label{Fig:EstError}
\end{figure}

Figure~\ref{Fig:EstError} shows the relative estimation errors ($||H^*[r]-H[r]||_2/||H[r]||_2$) of ADMOT for round $r\in \{1,2,..,50\}$.
 Note that we bound estimation error regardless the channel
stability $x\%$. Thus, lower channel stability only corresponds to more overheads (as shown in Figure~\ref{Fig:EstTime}).

\begin{figure}[htp]
\begin{center}
\includegraphics[height=87mm,width=97mm]{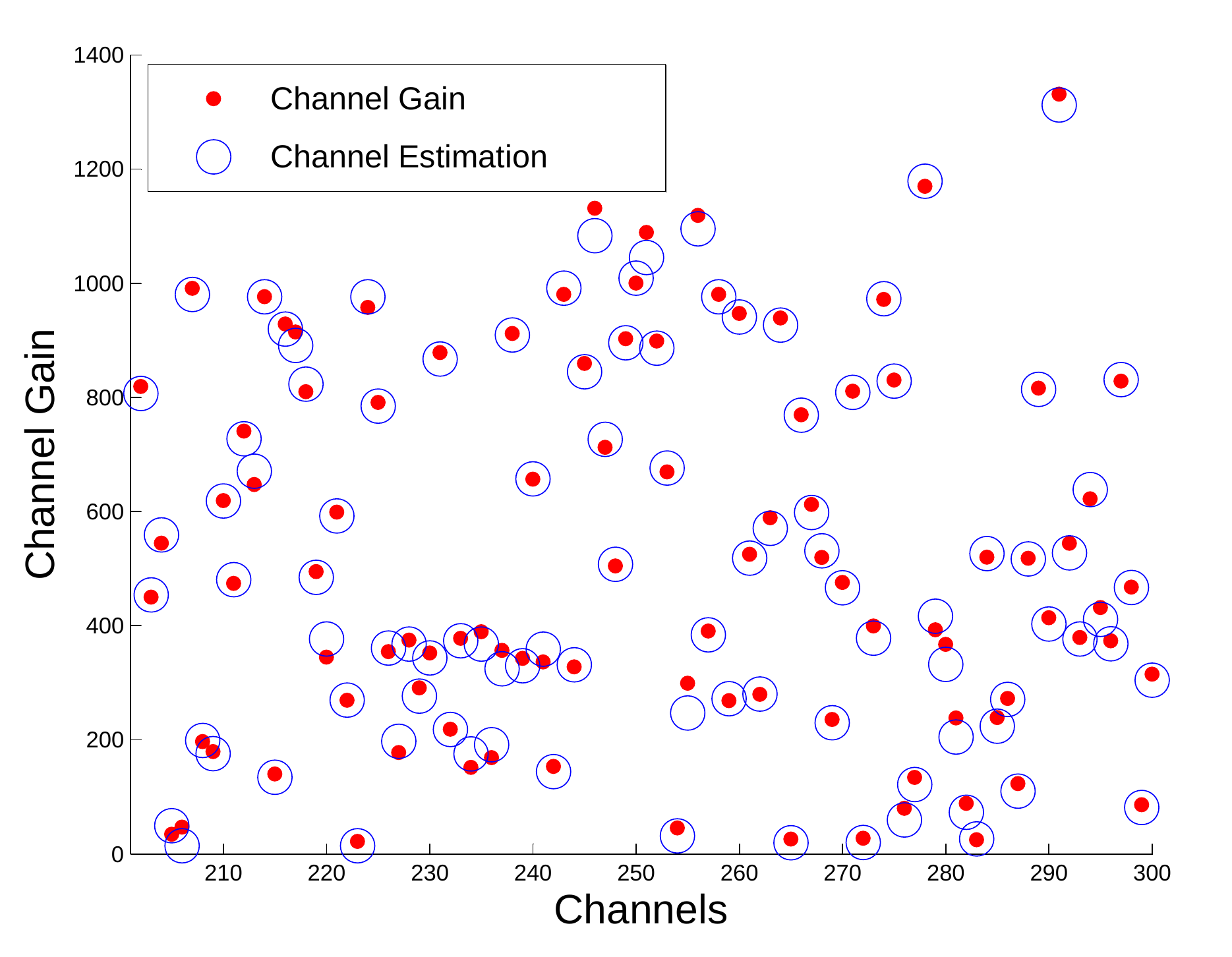}
 \end{center}
\caption{Detailed drawing of channel gains and the corresponding estimations.}
\label{Fig:EstDetail}
\end{figure}

For a detailed looking, we also show the estimations at round 50 for the case of $80\%$ channel stability.
Figure~\ref{Fig:EstDetail} draws (the absolute value of the real part) channel gains and the corresponding estimations for the $200,201,...,300$'th channels.

\section{Conclusion}
In the paper, we first investigate the scenario where a receiver needs to
track the channel gains of the channels with respect to $n$ transmitters. We assume that in current round of channel gain estimation,  no more than $k\leq n$ channels suffer significant variations since the last round.
We prove that ``$\Theta(k\log((n+1)/k))$ time slots'' is the {\it minimum} number of time slots needed to catch up with $k$ varying channels.
At the same time, we propose ADMOT to achieve the lower bound in a computationally efficient manner. Furthermore, ADMOT supports different {\it modulations} at the physical layer. Using above results, we also achieve the scaling law of general communication networks in which there are multiple transmitters, relay nodes and receivers. In the end of the paper, we also present simulation results to support our theoretical analysis.

\bibliographystyle{IEEEtran}
\bibliography{ADMOT}

\appendices

\section{Preliminaries of Compressive sensing \label{appendix: CS}}

Compressive sensing is a {\it mathematical} technique developed for compressible data
recovering with significantly fewer samples than
the length of data~\cite{CompressiveSensing1,CompressiveSensing2}.
All compressive sensing results used in the paper is introduced in this section.

Let $M$ be a matrix in $\Real^{m\times n}$ with $m\ll n$. Assume each column of $M$ is normalized to have $\ell_2$-norm $1$.
For positive integer $k$, $M$ is said to satisfy {\it restricted isometry property}(RIP) of order $k$ if
$(1-\delta_k)||X||^2_2\leq ||M X||^2_2\leq (1+\delta_k)||X||^2_2$
for all $k$-sparse vector $X\in \Real^n$~\cite{CompressiveSensingRobust}.

Let $A\in \Real^{m\times n}$ be a matrix. Then we have~\cite{SubGaussianRIP}:
\begin{itemize}
\item Each element of $A$ is i.i.d. generated from $\{-1,1\}$ with equal probability $\{1/2,1/2\}$. Then, with overwhelming probability ({\it i.e.}, $1-\mathcal O(2^{-n})$), matrix $A/\sqrt{m}$ satisfies RIP of order-$k$ provided that
$m\geq C_0 k \log{((n+1)/k)}$, where $C_0$ is a constant depending on each instance.

\item Each element of $A$ is i.i.d. generated from $\{0,-1,1\}$ with  probability $\{1/2,1/4,1/4\}$. Then, with overwhelming probability ({\it i.e.}, $1-\mathcal O(2^{-n})$), matrix $\sqrt{2/m}A$ satisfies RIP of order-$k$ provided that
$m\geq C_0' k \log{((n+1)/k)}$, where $C_0'$ is a constant depending on each instance.

\end{itemize}

Let $X\in \Real^n$ be the data vector and $Y=A X +Z$ be the noisy measurement,
 where $Z\in \Real^m$ is the noise with $||Z||_2\leq \sigma$. Let $X^*\in \Real^n$ be the solution
to ConvexOPT$(A,Y,\sigma)$, where ConvexOPT$(.)$ is defined in Section~\ref{subsec:consADMOT}.

Assuming $A/\sqrt{m}$ satisfies RIP, the following theorem is proved in~\cite{CompressiveSensingRobust}.
\begin{theorem}
\label{th:CS}
The solution $X^*$ obeys
\begin{equation}
||X-X^*||_2\leq C_1 d_k(X)/\sqrt{k}+C_2 \sigma/\sqrt{m},
\end{equation}
where $C_1$ and $C_2$ are constants and $d_k(.)$ is defined in~(\ref{eq:SparsityDistance}).
\end{theorem}

\section{Proof for Theorem~\ref{thm:ADMOT} \label{appendix: Proof}}

We show an intermediate lemma before the proof of Theorem~\ref{thm:ADMOT}.
Let $Z\in \Real^m$ such that each of its component $Z(i)$ is i.i.d. $\sim \mathcal N(0,1)$. Then we have

\begin{lemma}
\label{le:NormalNorm}
The $\ell_2$-norm $||Z||_2\leq \sqrt{2m}$ with a probability at least $1-e^{-0.15m}$.
\end{lemma}

\noindent {\bf Proof}:
 For any $i\neq j$, $Z(i)$ and $Z(j)$ are independent and normally distributed. Then the probability density function of $X=Z(i)^2+Z(j)^2$ is $f_X(x)=e^{-x/2}/2$ for $x\geq 0$~\cite{ProbabilityBook2005}. Thus,
$E\Big(e^{X/4}\Big)=\int_{0}^{+\infty}e^{(-x/4)}/2dx=2$.

Without loss of generality we assume ADMOT chooses $m$ as an {\it even} integer. Then we have:
\begin{eqnarray*}
& &Pr(||Z||_2^2>2m)=Pr\Big(\sum_{i=1}^d Z(i)^2/4>m/2\Big)\\
&=&Pr\Big(e^{\sum_{i=1}^m Z(i)^2/4}>e^{m/2}\Big)\\
&\leq& E\Big(e^{\sum_{i=1}^m Z(i)^2/4}\Big)\Big/ e^{m/2}\mbox{~~~~~~~~~~~~~~~~~~Markov Inequality }\\
&=& \prod_{j=1}^{m/2} E\Big(e^{Z(2j-1)^2/4+Z(2j)^2/4}\Big)\Big/e^{m/2} \mbox{~~~~~Independence}\\
&\leq&2^{m/2}/e^{m/2}\leq e^{-0.15m}.
\end{eqnarray*}
\hfill$\Box$

Then we have:

\noindent {\bf Proof of the Scaling laws in Theorem~\ref{thm:ADMOT}}:
Without loss of generality, we first consider a sub-problem ({\it i.e.}, an easier problem):
Assuming $Re(H-\hat H)$ is $(k,\sqrt{k} C)$-variation and $Im(H-\hat H)$ is a all-zero vector, what is the minimum time slots required to find
 $Re(H^*)$ such that $||Re(H^*)-Re(H)||_2^2=\mathcal O(1)$?

Assume $T$ time slots are used for  estimating $Re(H)$. For any
$s=1,2,...,T$, and $i=1,2,...,n$, in the $t$'th time slot let $S_i$ send $A(s,i)\in \Complex$. Here $A$ is a $T\times n$ complex
matrix whose $(s,i)$'th component is $A(s,i)$.

Let $Y(s)\in \Complex$ be the received data of $R$ in the $s$'th time slot, and therefore $Y\in \Complex^T$ be a length-$T$ vector whose $s$'th component is $Y(s)$.
Thus $Y=AH$. Note that we assume that there is no noise here, which only reduces the complexity of estimating $H$.

Since $\hat H$ and $A$ are known by $R$ as {\it a priori}, the original problem is equivalent to estimating
$\Delta=H-\hat H$ by $D=Y-A{\hat H}=A({H}-{\hat H})$. Due to $Im(H)=Im(\hat H)$, we have $Y=A(Re(H)-Re(\hat H))$. Thus the problem is equivalent to
estimating $Re(\Delta)$ by using $Re(D)$ and  $Im(D)$, which compose of $2T$ {\it linear samples} (over
$\Real$)
of $Re(\Delta)$.

A recent  result~\cite{CompressiveSensingLowerBound} in the field of compressive sensing proves that provided $d_k(Re(\Delta))\leq C \sqrt{k}$ for some constant $C$, it requires at least $\Theta(k\log((n+1)/k))$ linear samples (over $\Real$) for reliably finding $\Delta^*\in \Real^n$ such that
$||Re(\Delta)-\Delta^*||_2^2\leq \mathcal O(1)$. Thus we have $T\geq \Theta(k\log((n+1)/k))$.

Thus, we prove the complexity of the easier problem.  For the original problem, which considers random noise and the variations of imaginary parts
 of the channel gains, the complexity can only be higher.
\hfill$\Box$

\noindent {\bf Proof of the Achievability in Theorem~\ref{thm:ADMOT}}:
Recall that the constant $C_0$, $C_1$, and $C_2$ are defined in Appendix~\ref{appendix: CS}.
Note that since $m$ satisfies $m \geq C_0k\log((n+1)/k)$, with overwhelming probability ({\it i.e.}, $1-\mathcal O(2^{-n})$), the matrix $\Phi_m$ satisfies RIP of order-$k$ (Appendix~\ref{appendix: CS}). We henceforth assume it is true.

We first analyze $Re(D)=\Phi_m Re(H-\hat H)+Z$, where $Z\in \Real^{m}$ is the noise term. Using Lemma~\ref{le:NormalNorm}, we have $Pr(||Z||_2\leq \sqrt{2 m})<e^{-0.15m}$. We henceforth assume it is true.

Thus, using Theorem~\ref{th:CS}, vector $Re(\Delta*)$ satisfies
$||Re(\Delta^*)+Re(\hat H -H)||_2\leq C_1 d_k(Re(\hat H -H))/\sqrt{k}+\sqrt{2}C_2.$

Since the state variation $H-\hat H$ is $(k,\delta\sqrt{k})$-sparse, after setting $H^*=\hat H + \Delta^*$
we have $||Re(H^*-H)||_2\leq C_1 \delta + \sqrt{2}C_2$.

Similarly we have $||Im(H^*-H)||_2\leq C_1 \delta + \sqrt{2}C_2$. In the end, we have $||H^*-H||\leq \sqrt{2}C_1\delta+2C_2$.
It completes the proof.
\hfill$\Box$

\section{Proof of Theorem~\ref{le:D2Prob}\label{appendix: EstK}}

We first show an intermediate lemma.
Let $V\in \Real^n$ be a fixed vector, and $R\in \Real^n$ be a vector of random variables. For each component $R(i)$ we have $Pr(R(i)=1)=Pr(R(i)=-1)=0.5$, and $R(i)$ is i.i.d. for all $1\leq i\leq n$.
Let $<V,R>=\sum_{i=1}^n V(i)R(i)$ denote the inner product between $V$ and $R$.
\begin{lemma}
\label{le:L2Inner}
For $<V,R>^2$, the expectation $E\Big(<V,R>^2\Big)$ is $||V||_2^2$ and the variance Var$(<V,R>^2)$ is no more than $||V||_2^4$.
\end{lemma}
\noindent{\bf Proof}:
We have
\begin{eqnarray*}
<V,R>^2&=&\sum_i R(i)^2V(i)^2+\sum_{i\neq j} R(i)R(j)V(i)V(j)\\
&=&V(i)^2+\sum_{i\neq j} R(i)R(j)V(i)V(j)\\
&=&||V||_2^2+\sum_{i\neq j} R(i)R(j)V(i)V(j).
\end{eqnarray*}
Since $E(\sum_{i\neq j} R(i)R(j)V(i)V(j))=0$, we have $E\Big(<V,R>^2\Big)=||V||_2^2$.

We have:
\begin{eqnarray*}
E\Big(<V,R>^4\Big)&=&||V||_2^4+E\Big(\Big(\sum_{i\neq j} R(i)R(j)V(i)V(j)\Big)^2\Big)\\
&=&||V||_2^4+E\Big(\Big(\sum_{i\neq j} R(j)^2R(i)^2V(j)^2V(i)^2\Big)\Big)\\
&=&||V||_2^4+\Big(\sum_{i\neq j} V(j)^2V(i)^2\Big)\\
&\leq& ||V||_2^4+\Big(\sum_i V(i)^2\Big)\Big(\sum_j V(j)^2\Big)\\
&=&2||V||_2^4.
\end{eqnarray*}
Thus  Var$\Big(<V,R>^2\Big)=E\Big(<V,R>^4\Big)-E\Big(<V,R>^2\Big)^2$ is no more than $||V||_2^4$.
\hfill$\Box$

Then we have:

\noindent{\bf Proof of Theorem~\ref{le:D2Prob}}:
Let $\Delta=(H-H_1^*)$, $\varphi_R=||Re(\Delta)||_2$,
$\varphi_I=||Im(\Delta)||_2$ and thus $\varphi=||\Delta||_2=\sqrt{\varphi_R^2+\varphi_I^2}$.

Without loss of generality, we first analyze  $||Re(D_2)||_2^2$.

From the definition, $Re(D_2)=U_R+Z_R$, where each component of $Z_R\in \Real^d$ is i.i.d.$\sim\mathcal N(0,1)$  and $U_R\in \Real^d$ is $\Phi_{(m,2)}\Delta_R$. Using Lemma~\ref{le:L2Inner}, for $i=1,2,...d$, we have $E(U_R(i)^2)=\varphi_R^2$ and Var$(U_R(i)^2)\leq \varphi_R^4$.
Note that each component of $\Phi$ is i.i.d.; the noise term is i.i.d. $\sim \mathcal N(0,1)$; $U_R(i)$ and $U_R(j)$ are independent for all $i\neq j$. Thus, we can apply Chernoff Bound (on discrete bounded random variables)~\cite{ChernoofDiscrete} and get
$Pr\Big(|\sum_{i=1}^d(U_R(i)^2-\varphi_R^2)/(n\varphi_R)|>d\varphi_R^2/(2n\varphi_R)\Big)\leq 2e^{-d^2/16}$.
It is equivalent to $$Pr(d\varphi_R^2/2\leq ||U_R||_2^2\leq 3d\varphi^2_R/2)\geq 1- 2e^{-d^2/16}$$.

Using Lemma~\ref{le:NormalNorm}, we have $Pr(||Z_R||_2^2>2d)\leq e^{-0.15d}$.

Similarly, assuming $Im(D_2)=U_I+Z_I$, we have
\begin{eqnarray*}
& &Pr\Big(d\varphi_I^2/2\leq ||U_I||_2^2\leq 3d\varphi^2_I/2\Big)\geq 1- 2e^{-d^2/16},\\
& &Pr(||Z_I||_2^2>2d)\leq e^{-0.15d}.
\end{eqnarray*}

Using Union Bound~\cite{ProbabilityBook2005}  for $U_R$ and $U_I$, we have
\begin{eqnarray*}
Pr(d\varphi^2/2\leq ||U_R+jU_I||_2^2\leq 3d\varphi^2/2)\geq 1- 4e^{-d^2/16}.
\end{eqnarray*}

Similarly, by the Union Bound for $Z_R$ and $Z_I$ we can derive
\begin{eqnarray*}
Pr(||Z_R+jZ_I||_2^2>4d)\leq 2e^{-0.15d}.
\end{eqnarray*}

Note that $D_2=(U_R+jU_I)+(Z_R+jZ_I)$.  Using {\it triangle inequality}, the event $||D_2||_2^2>d(\varphi\sqrt{3/2}+2)^2)$ happens with
a probability at most $4e^{-d^2/16}+2e^{-0.15d}=O\Big(e^{-0.15d}\Big)$.

Assuming $\varphi^2>8$, the event $||D_2||_2^2<d(\varphi/\sqrt{2}-2)^2$  happens with
a probability at most $4e^{-d^2/16}+2e^{-0.15d}=O\Big(e^{-0.15d}\Big)$.
\hfill$\Box$

\end{document}